\documentclass[a4paper,11pt]{article}
\usepackage{pos}
\usepackage{amsmath,graphicx,multirow,tabularx,physics}
\usepackage{davitex,slashed,listings}
\usepackage{lineno}
\newcommand \beqa {\begin{eqnarray}}
\newcommand \eeqa {\end{eqnarray}}
\newcommand \hmu {\hat{\mu}}
\newcommand \hs {\hat{s}}

\newcommand \he {\hat{\epsilon}}
\newcommand \hp {\hat{p}}

\newcommand \hn {\hat{n}}
\newcommand{\pade}{Pad\'e~}
\newcommand \xb {\bar{x}}
\usepackage{amsmath,graphicx,multirow,tabularx,physics}

\title{The isentropic equation of state of (2+1)-flavor QCD: An update based on high precision Taylor expansion and Pad\'e-resummed expansion at finite chemical potentials}
\ShortTitle{Isentropic equation of state of (2+1)-flavor QCD}

 \manuallySeparateAuthors
\author*[a]{Jishnu Goswami}
\author{(HotQCD collaboration)}

\affiliation[a]{RIKEN Center for Computational Science,\\ 
Kobe 650-0047, Japan}
\emailAdd{jishnu.goswami@riken.jp}

\abstract{
The HotQCD Collaboration performed Taylor expansion calculations in 2017 for the pressure, energy density, and entropy density at non-zero chemical potentials up to the $6\nth$ order. Since then, they have significantly improved the statistics for lattices with temporal extents of $N_\tau=8$ and $12$, and have also included results for $N_\tau=16$ that were not previously available. They have also calculated the $8\nth$-order expansion coefficients for $N_\tau=8$. These calculations showed that the Taylor series expansion for the pressure is accurate up to $\mu_B / T \leq 2.5$. In this study, we use the high-statistics results on Taylor expansion coefficients, calculated with HISQ fermions and extrapolated to the continuum limit, to determine the QCD equation of state under conditions relevant for hot and dense matter produced in heavy ion collisions. We also calculate the energy density and pressure along lines of constant entropy per net baryon number.
}

\FullConference{%
 The 39th International Symposium on Lattice Field Theory, LATTICE2022
  26th-30th July, 2022
  Bonn, Germany
}

\begin{document}
\maketitle

\section{Introduction}\label{sec:intro}
Quantum chromodynamics (QCD) is studied at finite temperature and chemical potential in order to understand the properties of hot and dense matter created in heavy ion collision (HIC) experiments. To describe the state of the matter at extreme conditions, the equation of state (EoS) of QCD in the grand canonical ensemble at finite temperature and non-zero chemical potentials is important. This EoS is an essential input for interpreting heavy ion data \cite{Braun-Munzinger:2015hba} in thermal equilibrium and for hydrodynamic modeling of matter created in HIC experiments. It is also used in the analysis of the ``cosmic trajectory" \cite{Middeldorf-Wygas:2020glx} at high temperature and low chemical potentials, and for the EoS of neutron star mergers \cite{Drischler:2020fvz,Fujimoto:2022ohj} at low temperature and high chemical potentials.

The EoS of (2+1)-flavor QCD, or the relationship between the pressure, energy density, entropy density and temperature of a system, has been well studied at vanishing chemical potential using lattice techniques in \cite{Karsch:2000ps, HotQCD:2014kol,Borsanyi:2013bia}. However, it is difficult to calculate the EoS at non-vanishing chemical potentials due to the sign problem. Currently, the most commonly used methods for determining the EoS at non-vanishing chemical potentials are Taylor expansion and analytic continuation, both of which have their own strengths and limitations. Taylor expansion, analytic continuation and various resummation technique have been used recently to analyze the EoS at non-vanishing values of the chemical potentials for baryon number, electric charge, and strangeness in previous lattice QCD calculations \cite{Gavai:2004sd,Mitra:2022vtf,Mitra:2022zsa,Borsanyi:2022qlh,Schmidt:2021pey, Nicotra:2021ijp,Dimopoulos:2021vrk}. 

Recently, we proposed the use of the \pade approximation~\cite{Bollweg:2022rps} to improve upon ordinary Taylor
expansions of pressure, which we hope will also increase the reliability of the EoS at finite chemical potentials. We have updated our previous analysis of the EoS of (2+1)-flavor QCD using a large data set of gauge field configurations generated using \texttt{SIMULATeQCD}~\cite{Bollweg:2021cvl} with the HISQ action. A new data set for $N_\tau = 16$ and a larger data set for temperatures between $125$-$175$ MeV were used to update our analysis of the EoS in (2+1)-flavor QCD using the HISQ/tree action. The increased statistics, which include approximately 1.5 million configurations on $32^3\times 8$ lattices at the pseudo-critical temperature of 156.5(1.5) MeV, allowed us to extend our previous Taylor series results to eighth-order in the chemical potentials. We use the publicly available \texttt{AnalysisToolbox}~\cite{toolbox} for most of the data analysis in this proceeding.
 \begin{figure*}[htbp]
\centering
\includegraphics[scale=0.54]{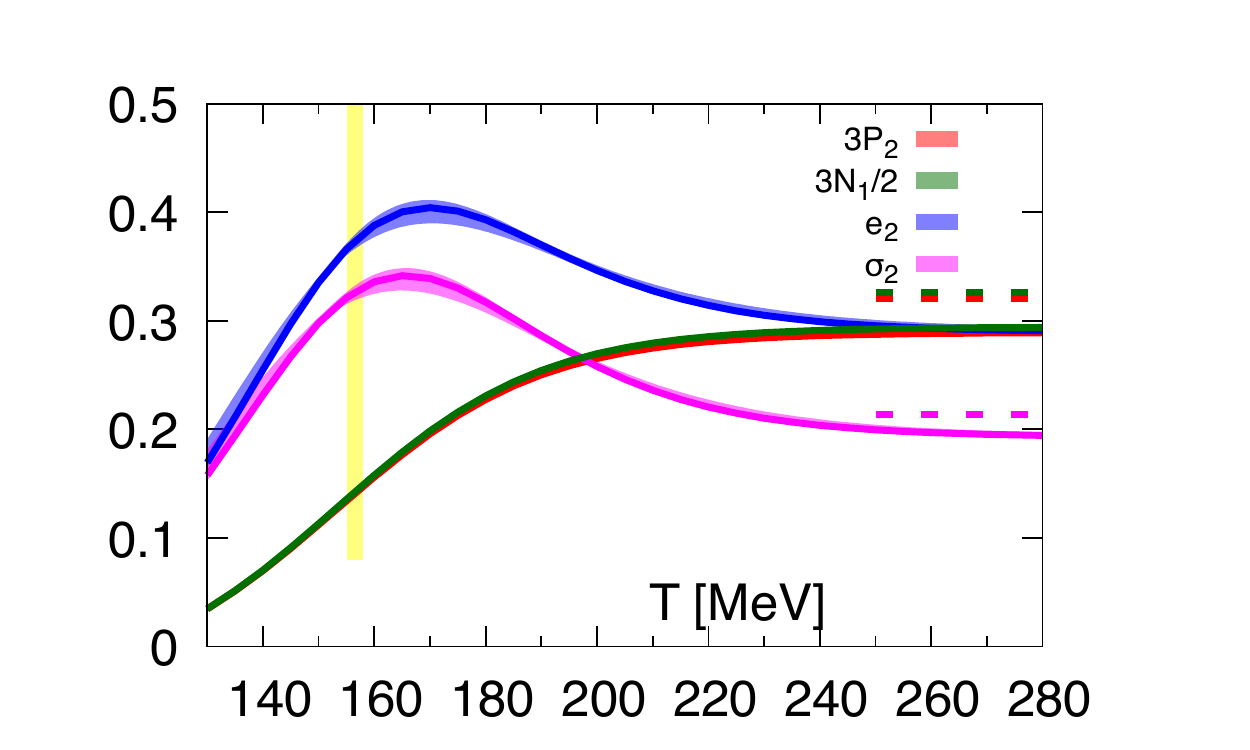}
\includegraphics[scale=0.54]{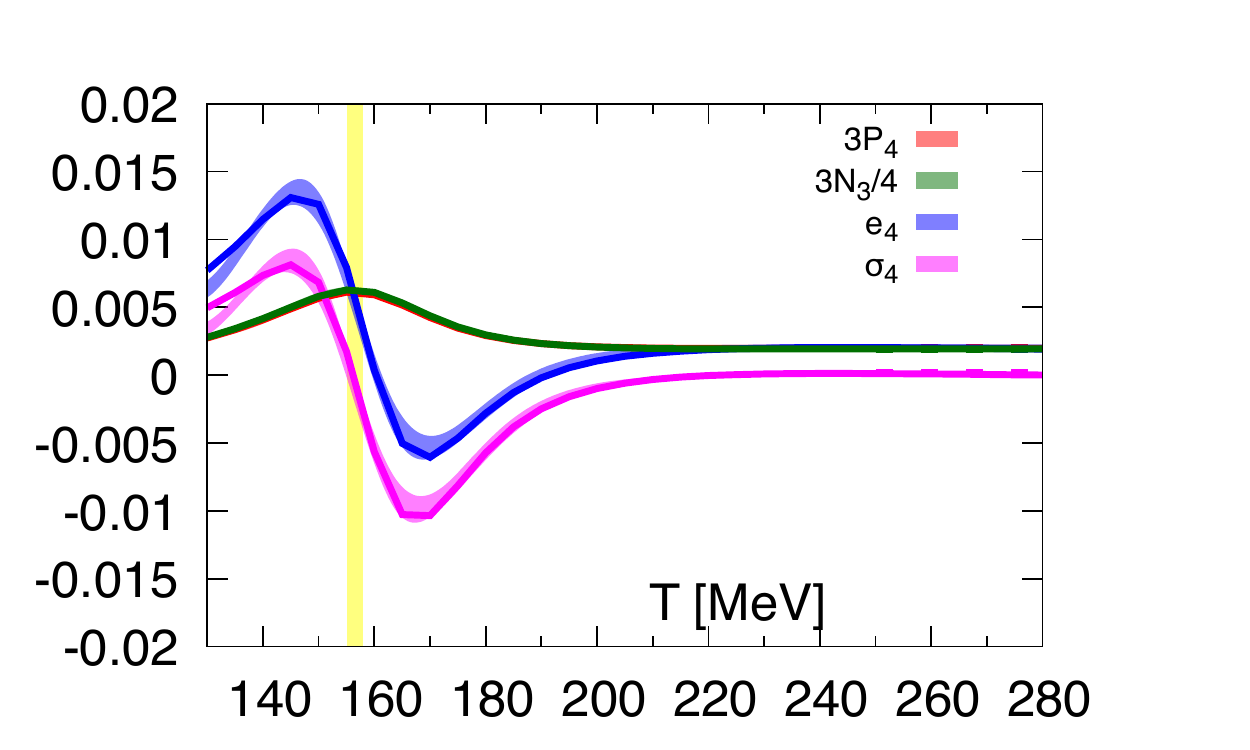}
\includegraphics[scale=0.54]{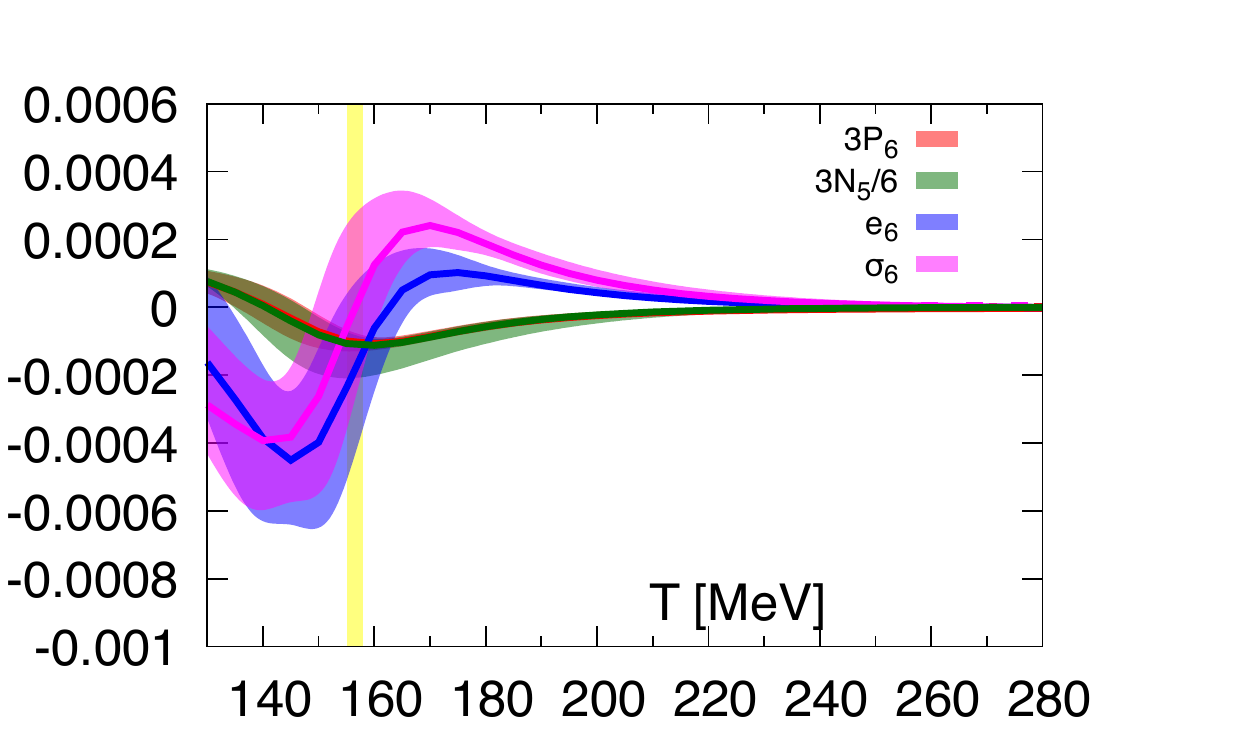}
\includegraphics[scale=0.54]{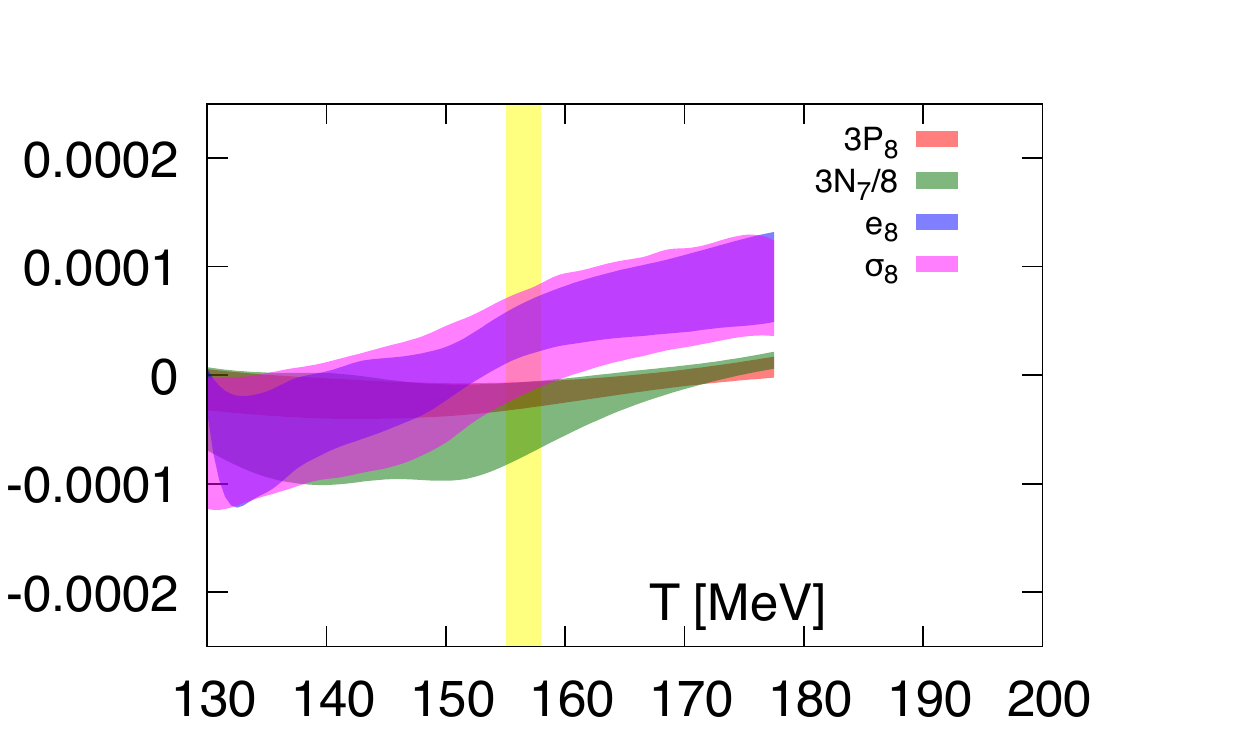}
\caption{Second-order (top, left), fourth-order (top, right), sixth-order (bottom, left) and eighth-order (bottom, right) expansion coefficients of pressure ($P_{2k}$), net baryon number density ($N_{2k-1}$), energy density ($\epsilon_{2k}$) and entropy density
($\sigma_{2k}$) for a strangeness-neutral 
medium ($n_S=0$) with electric charge to 
baryon-number density $n_Q/n_B = r= 0.4$. The solid lines are constructed from the parametrization in Table \ref{tab:pfit}.}
\label{fig:exp_coeff}
\end{figure*}

\section{Taylor expansion and \pade resummation of bulk thermodynamics}
The notation used for the Taylor expansions of bulk thermodynamic observables in $(2+1)$-flavor QCD is outlined in \cite{Bazavov:2017dus}. For ease of reference, we provide a brief summary below,
\begin{eqnarray}
	\frac{P}{T^4} = \hat{p} &=& \frac{1}{VT^3}\ln Z(T,V,\vec{\mu}) = \sum_{i,j,k=0}^\infty%
\frac{\chi_{ijk}^{BQS}}{i!j!\,k!} \hmu_B^i \hmu_Q^j \hmu_S^k \; , \\
\hat{\epsilon} &=&\frac{1}{VT^3} T\frac{\partial \ln Z(T,V,\hmu_B,\hmu_Q,\hmu_S)}{\partial T} 
= \sum_{i,j,k=0}^\infty \frac{\Xi_{ijk}^{BQS} + 3\chi_{ijk}^{BQS}}{i!j!\,k!} \hmu_B^i \hmu_Q^j \hmu_S^k\; ,  \\
\hs &=& 
\he +\hp -\hmu_B \hn_B - \hmu_Q \hn_Q -\hmu_S \hn_S \; .
\label{Pdefinition}
\end{eqnarray}
with $\hat{\mu}_X\equiv \mu_X/T$.
$\chi_{ijk}^{BQS}$ and $\Xi_{ijk}^{BQS}$ can be written as,
\begin{equation}
\chi_{ijk}^{BQS} =\left. 
\frac{1}{VT^3}\frac{\partial \ln Z(T,V,\vec{\mu}) }{\partial\hmu_B^i \partial\hmu_Q^j \partial\hmu_S^k}\right|_{\vec{\mu}=0} \; \; ,  \; \Xi_{ijk}^{BQS} = \frac{Td\chi_{ijk}^{BQS}}{dT} ,
\; i+j+k\; {\rm even} \; .
\label{suscept}
\end{equation}

We will study matter that is strangeness neutral ($n_S=0$) at a fixed ratio of electric charge to baryon number ($n_Q/n_B=r$) 
by introducing constraints on the electric charge and strangeness chemical potentials.
This makes $\hmu_Q$ and $\hmu_S$ functions of $T$ and $\hmu_B$, and hence we can expand
\begin{eqnarray}
        \hat{\mu}_S(T,\hmu_B) &=& s_1(T)\hat{\mu}_B + s_3(T) \hat{\mu}_B^3  + s_5(T) \hat{\mu}_B^5 + s_7(T) \hat{\mu}_B^7 \nonumber \\
         \hat{\mu}_Q(T,\hmu_B) &=& q_1(T)\hat{\mu}_B + q_3(T) \hat{\mu}_B^3 + q_5(T) \hat{\mu}_B^5 + q_7(T) \hat{\mu}_B^7.
\label{qs}
\end{eqnarray}
The explicit expressions of $q_i$ and $s_i$ for $i=1, 3, 5, 7$ can be found in \cite{HotQCD:2017qwq,Bazavov:2020bjn}. By substituting the expressions of $\mu_Q$ and $\mu_S$ using Eq.~\ref{qs}, one can calculate the Taylor series expansion for the number density, the $\hmu_B$-dependent part of the pressure, the energy density, and the entropy density,
\begin{eqnarray}
\frac{n_B}{T^3} &=& \sum_{k=1}^{\infty} N_{2k-1}(T) \hmu_B^{2k-1} \\
\frac{\Delta \mathcal{O}}{T^4}=\frac{\mathcal{O}(T,\mu_B) - \mathcal{O}(T,0)}{T^4} &=& \sum_{k=1}^{\infty} \mathcal{O}_{2k}(T) \hmu_B^{2k} \ , \text{~where}~~\mathcal{O} = P,\epsilon, \\
\frac{\Delta s}{T^3}=\frac{s(T,\mu_B) - s(T,0)}{T^3} &=& \sum_{k=1}^{\infty} \sigma_{2k}(T) \hmu_B^{2k}.
\label{chineutral}
\end{eqnarray}
The expansion coefficient of the number density and electric charge chemical potentials will be related to other observables in the following way,
\begin{eqnarray}
P_{2n} &=& \frac{1}{2n}\left(N_{2n-1}^B + r\sum\limits_{k=1}^{n}(2k-1)q_{2k-1}N^B_{2n-2k+1}\right) \label{eq:p} \\
P^{\prime}_{2n} &=& \frac{1}{2n}\left(N_{2n-1}^{B^{\prime}} + r\sum\limits_{k=1}^{n}(2k-1)\left(q_{2k-1}^{\prime}N^B_{2n-2k+1}+q_{2k-1}N^{B^\prime}_{2n-2k+1}\right)\right) \label{eq:dp}\\
\epsilon_{2n} &=& 3P_{2n} + TP^{\prime}_{2n} - r \sum\limits_{k=1}^{n}Tq^{\prime}_{2k-1}N^B_{2n-2k+1} \label{eq:e} \\
\sigma_{2n} &=& 4P_{2n} + TP^{\prime}_{2n} - N_{2n-1}^B - r \sum\limits_{k=1}^{n}\left(q_{2k-1} + Tq^{\prime}_{2k-1}\right)N^B_{2n-2k+1}. \label{eq:s}
\end{eqnarray}

In Fig.~\ref{fig:exp_coeff}, we present the results for the expansion coefficients of the pressure series, as well as those for the number density, energy density and entropy density for $r = 0.4$. For $n=1$ and $2$, the coefficients for the energy density and entropy density have been extrapolated to the continuum limit. For $n=3$ and $4$, we provide spline interpolations of the results obtained using $N_\tau=8$.

\begin{figure*}[htbp]
\centering
\includegraphics[width=0.50\textwidth]{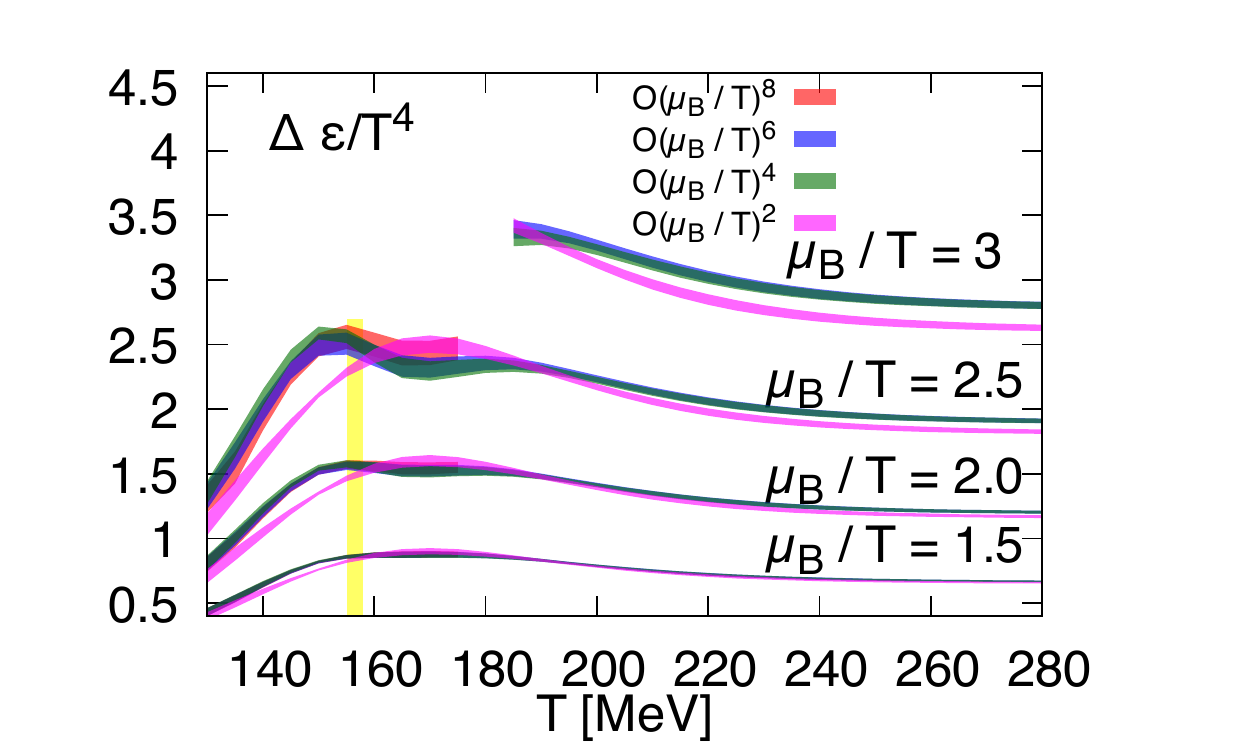}\hspace{-6mm}
\includegraphics[width=0.50\textwidth]{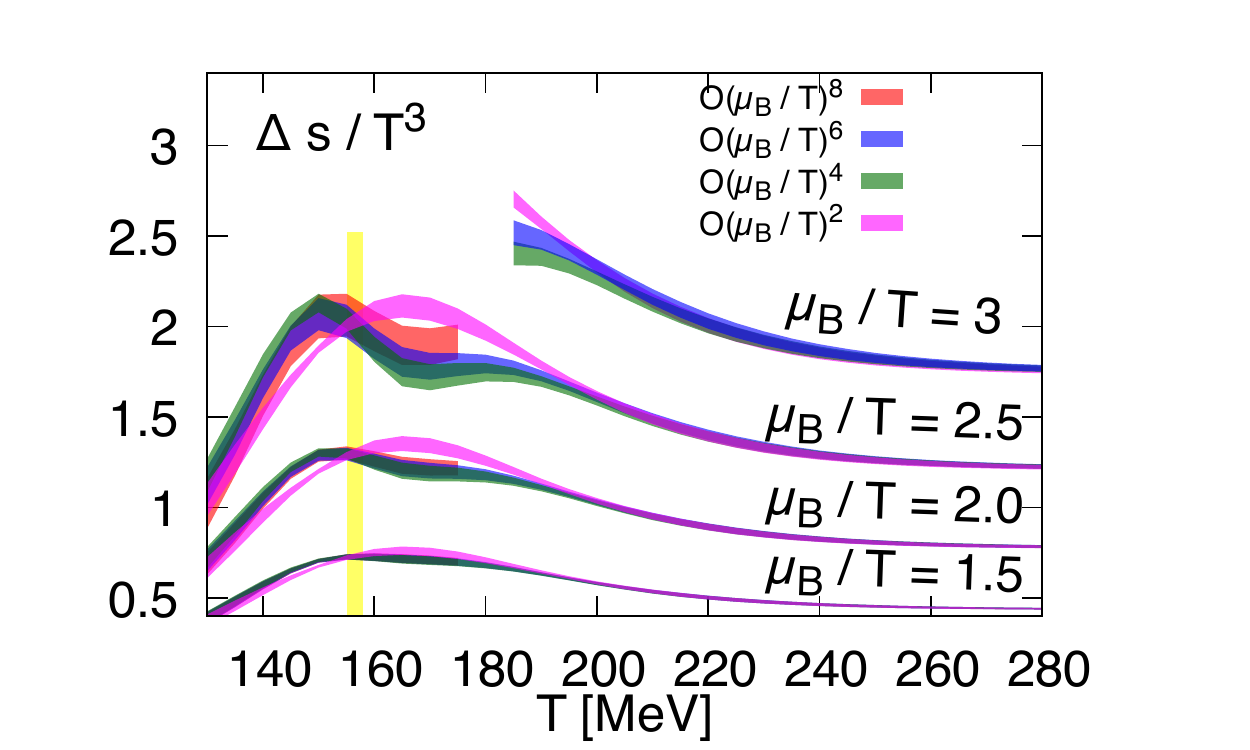}
\includegraphics[width=0.50\textwidth]{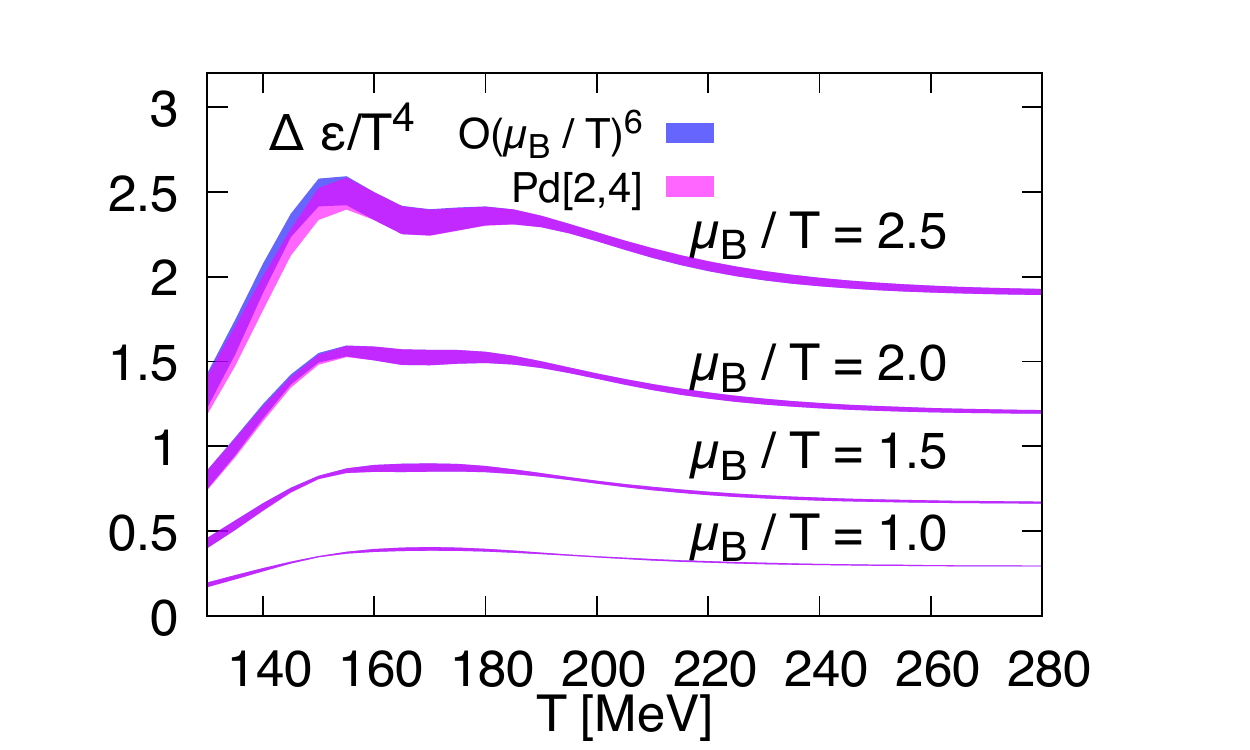}\hspace{-6mm}
\includegraphics[width=0.50\textwidth]{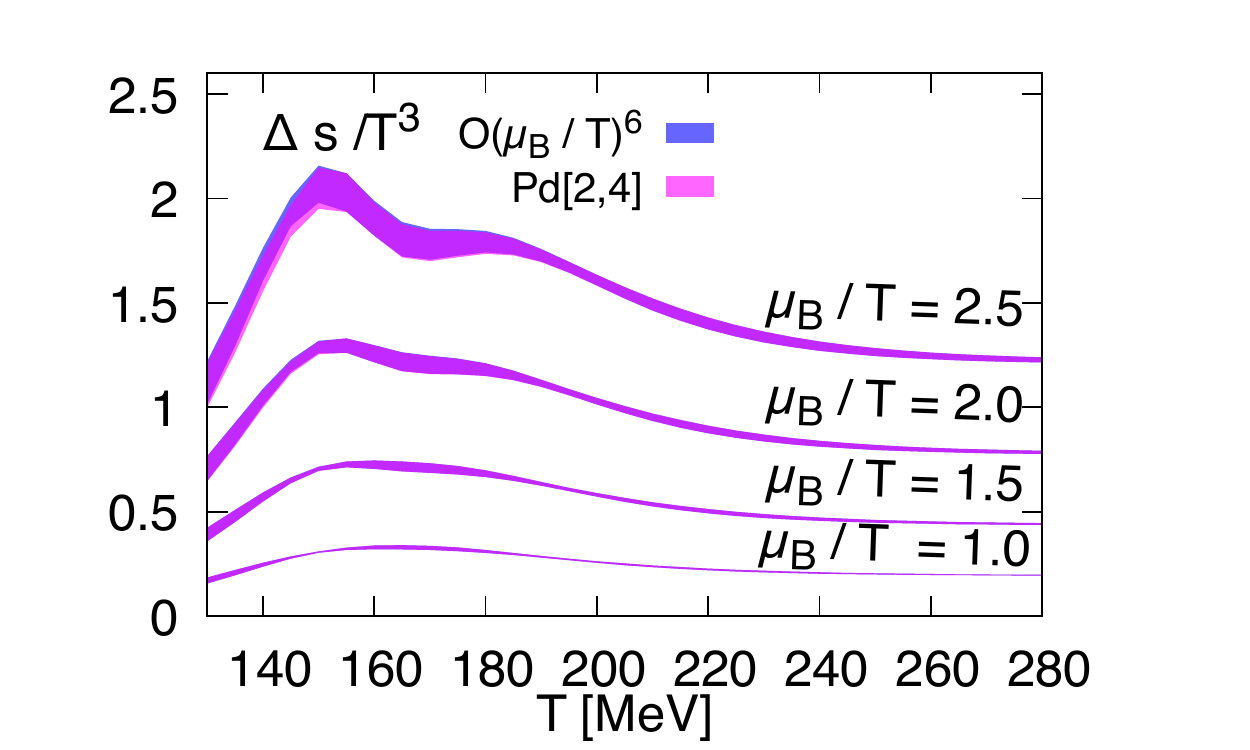}
\caption{Comparison of different (top) order Taylor expansion
results for $\Delta \epsilon/T^4$ with corresponding [2,4] \pade approximants. 
The yellow bands highlight the $\Tpc=156.5(1.5)$ MeV.
}
\label{fig:del-e}
\end{figure*}

In order to understand the range of validity of energy density and entropy density as a 
function of baryon chemical potential, we will use Taylor series and \pade approximants to compare different orders. 
Our focus will be on strangeness-neutral matter with a ratio of electric charge density to baryon number density of $r = 0.4$, 
which is seen in heavy ion collision experiments where the net strangeness number density is zero
and the ratio of net electric charge to net baryon number density is $0.4$ \cite{Bazavov:2017dus}. 
As part of our analysis, we will define the [2,4] \pade approximant for energy and entropy densities,
\begin{eqnarray}
    \left( \frac{\Delta \epsilon(T,\mu_B)}{T^4}\right)_{[2,4]} &=&\frac{\epsilon_2^2}{\epsilon_4} \frac{\xb^2}{1-\xb^2 + (1-c_{6,2}) \xb^4} \; , \xb =\sqrt{\epsilon_4/\epsilon_2}\ \hmu_B \;, c_{6,2}=\frac{\epsilon_6 \epsilon_2}{\epsilon_4^2} .  \label{energyPade} \\
       \left( \frac{\Delta s(T,\mu_B)}{T^3}\right)_{[2,4]} &=&\frac{\sigma_2^2}{\sigma_4} \frac{\xb^2}{1-\xb^2 + (1-c_{6,2}) \xb^4} \; , \xb =\sqrt{\sigma_4/\sigma_2}\ \hmu_B \;, c_{6,2}=\frac{\sigma_6 \sigma_2}{\sigma_4^2} .  \label{entropyPade}
\end{eqnarray} 

In Fig.~\ref{fig:del-e}, we present the energy density (top, left) and entropy density (top, right) for various values of $\hmu_B$ as a function of temperature. We have found that for high temperatures ($T\gtrsim200$MeV), the Taylor series quickly converges for these observables, as reported in \cite{HotQCD_Eos}. We also show results for larger chemical potentials, such as $\hmu_B\simeq3$, for energy and entropy density at temperatures above $200$ MeV. In Fig.~\ref{fig:del-e} (bottom, left), we compare the results of the $6\nth$-order Taylor series for $\Delta \epsilon$ with the corresponding [2,4] \pade approximants introduced in Eq.\ref{energyPade}. We see that the Taylor series and the [2,4] \pade approximants agree well up to a baryon chemical potential of $\mu_B/T \gtrsim 2.5$. However, at the highest chemical potential, the energy density becomes somewhat ``wiggly." The reliability of the energy density is therefore limited to $2<\mu_B/T<2.5$. Similar conclusions can be drawn from the comparison of the Taylor expansion and \pade approximant of the entropy density show in Fig.~\ref{fig:del-e} (bottom, right).

\begin{figure*}
\centering
\includegraphics[width=0.35\textwidth]{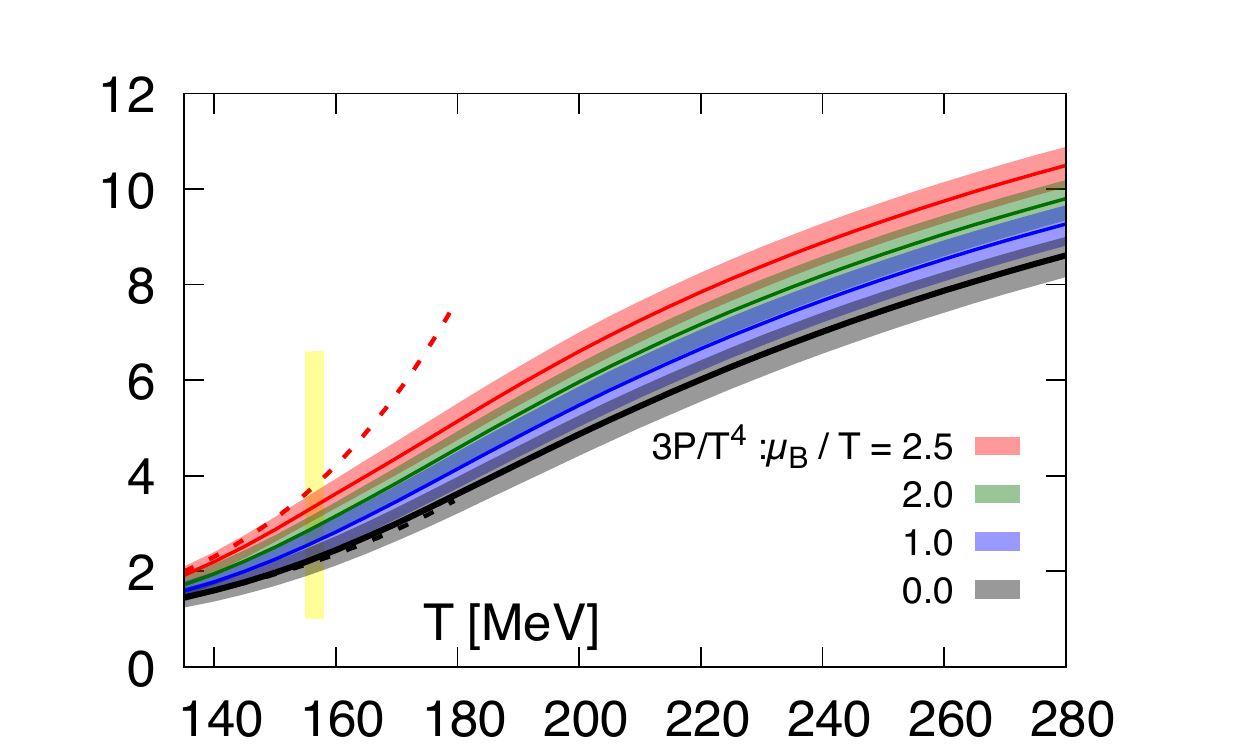}\hspace{-6mm}
\includegraphics[width=0.35\textwidth]{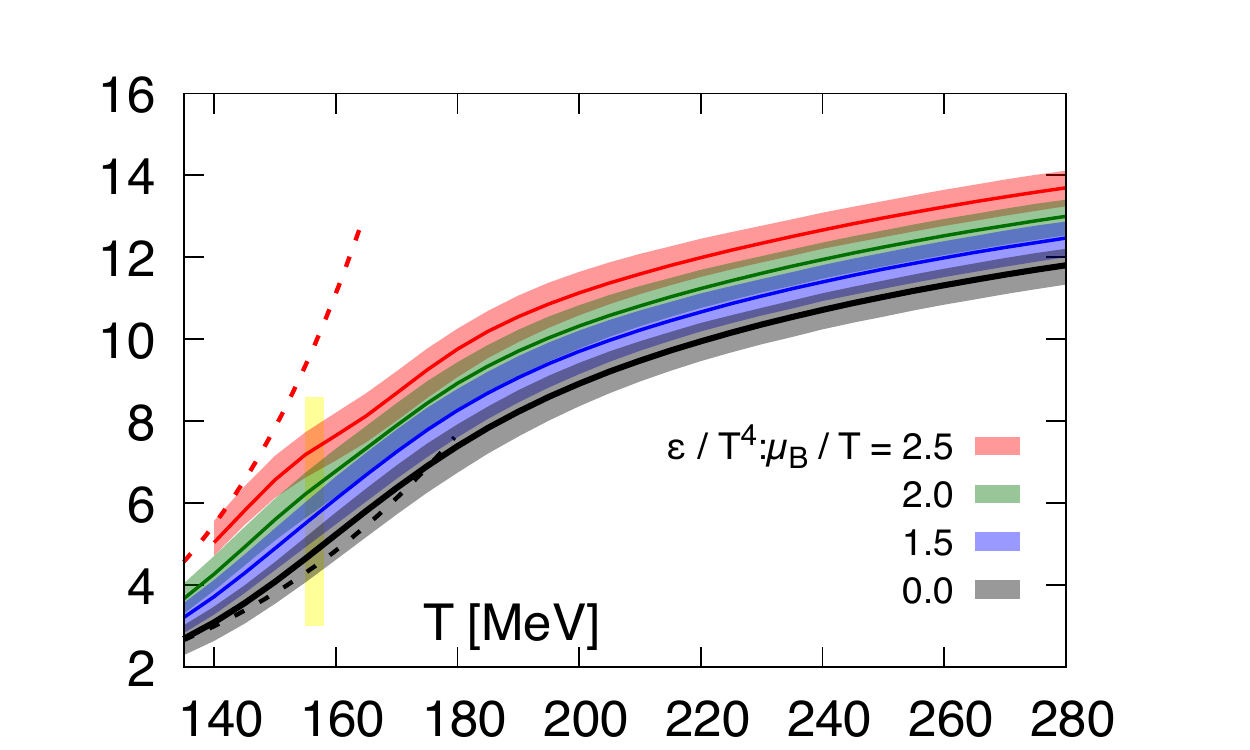}\hspace{-6mm}
\includegraphics[width=0.35\textwidth]{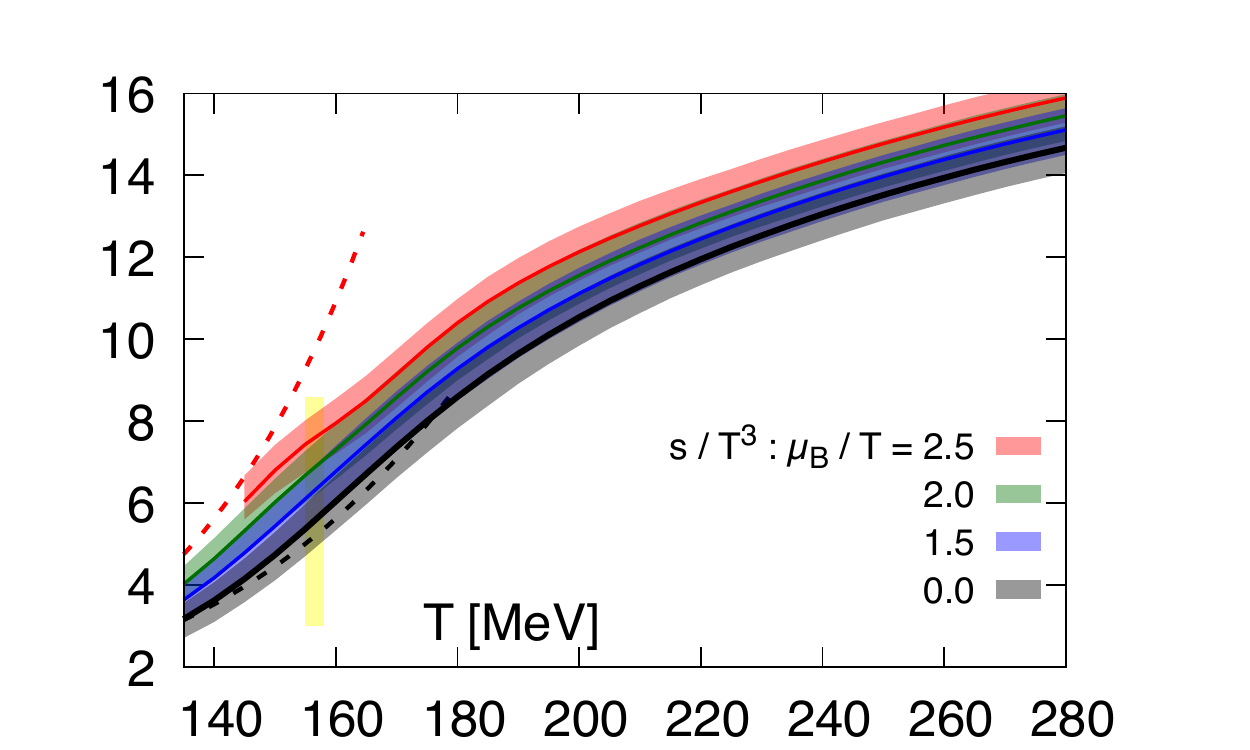}
\caption{Pressure (left), energy (middle) and entropy (right) densities versus temperature
for several values of the baryon chemical potential. Figures 
show results for the case $n_S=0$,
$n_Q/n_B=0.4$ in the temperature interval [130 MeV:280 MeV]. The results and parametrization for $\hmu_B=0$ were taken from \cite{HotQCD:2014kol}. The solid lines are produced using the parametrization listed in Table \ref{tab:pfit}, while the dotted lines are based on the QMHRG2020~\cite{Bollweg:2021vqf,Goswami:2021opr} calculation.
}
\label{fig:pes-fixedmu}
\end{figure*}

\begin{figure*}
\includegraphics[scale=0.63]{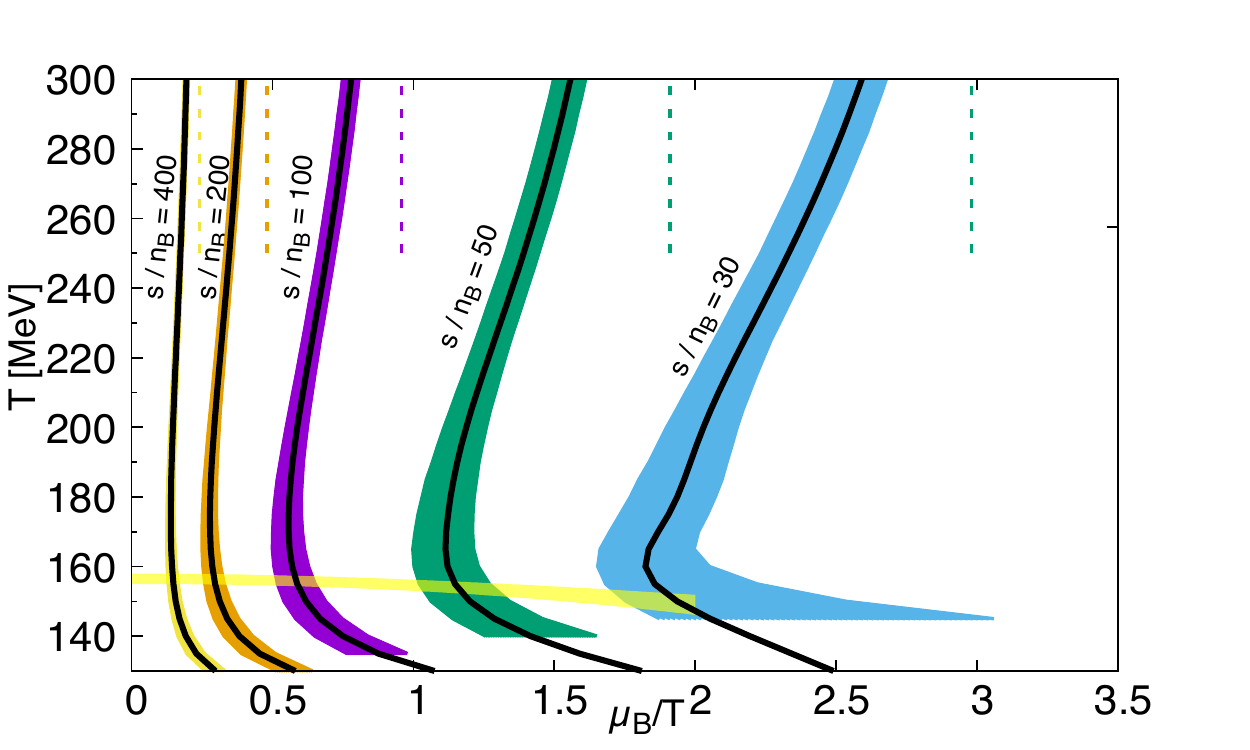}\hspace{-6mm}
\includegraphics[scale=0.63]{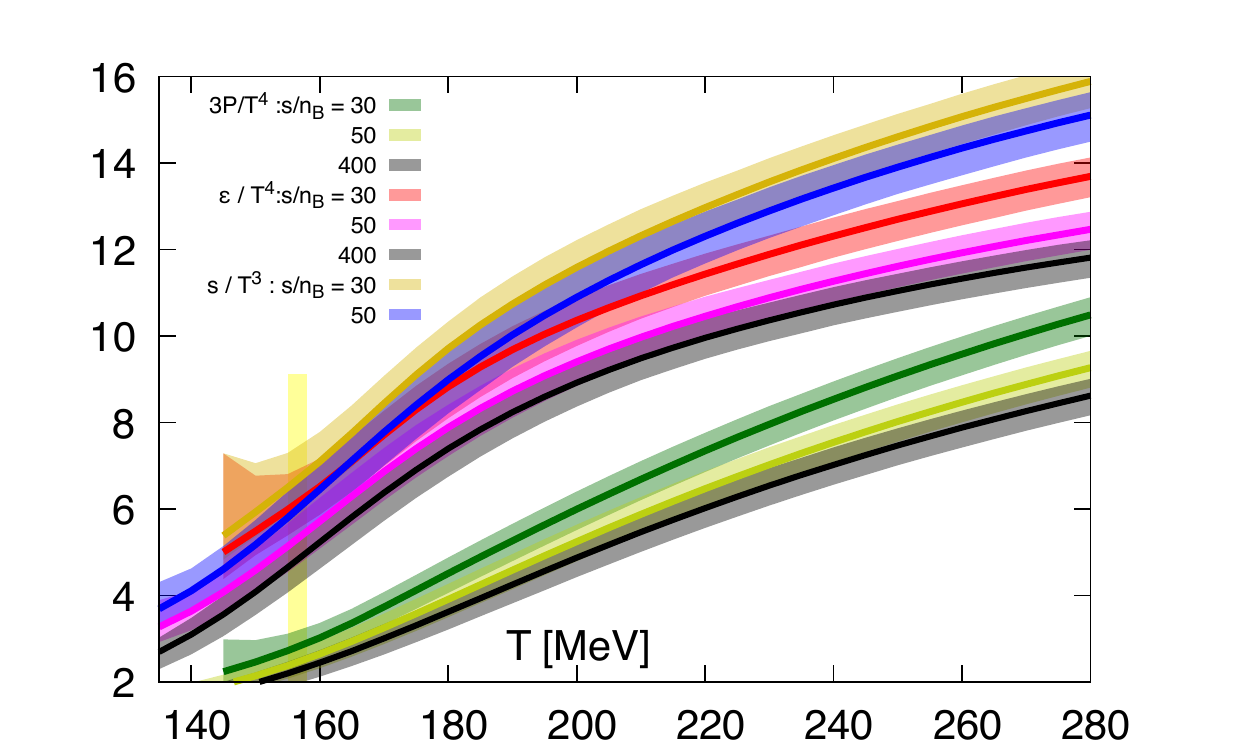}
\caption{Lines of constant entropy per baryon number (left) in the $T$-$\hmu_B$
plane. Pressure, energy and entropy densities (right) versus temperature
in constant $s/n_B$ trajectories.}
\label{fig:pes-fxiedsB}
\end{figure*}

\section{Parametrization of the EoS of (2+1)-flavor QCD}
In Fig.~\ref{fig:pes-fixedmu}, we present the total pressure (left), energy (middle), and entropy (right) 
densities for $\hmu_B \in [0:2.0]$ at all temperatures analyzed. For $\hmu_B=2.5$, 
we only show results for energy and entropy density at temperatures $T\ge 140$~MeV. 
The EoS at fixed $\{n_S=0,n_Q/n_B=0.4,\mu_B/T\}$ is represented by these results. In this section, 
we will also also update the parametrization for $\mu_B > 0$ from \cite{Bazavov:2017dus}.
The $N_k$'s and $q_k$'s (where $k=1,3,5$) can be used to create all expansion coefficients of the bulk thermodynamic observables, as shown by Eq.~(\ref{eq:p}-\ref{eq:s}). 
The functional forms of the $N_k$'s and $q_k$'s are defined as follows:

\begin{table}[t]
	\scriptsize
	\begin{tabular}{|c||c|c|c|c|c|c|c|c|c|} \hline\hline
		$k$&$N^B_{k,0n}$&$N^B_{k,1n}$&$N^B_{k,2n}$&$N^B_{k,3n}$&$N^B_{k,4n}$ &
		$N^B_{k,1d}$&$N^B_{k,2d}$&$N^B_{k,3d}$&$N^B_{k,4d}$ \\
		\hline
		1&0.181146&-0.42891324& 0.40657622 &-0.14809017 & 0.00584306 &-2.79490041& 3.91094903 & -3.26388671 &1.33731772 \\
		3& -0.00553136 & 0.04533642 & -0.09913670  & 0.08379059 &-0.02394836 & -0.17023948 & -1.82763726 & 0.0 & 1.05942220 \\
		5& 1.3211e-04 & -9.0274e-04 & 0.00227410 & -0.00247477 & 9.5899e-04 & -1.06474394 & 0.19724813 & -1.02866873 &  0.95555405 \\
		\hline
		\hline
		$k$&$q_{k,0n}$&$q_{k,1n}$&$q_{k,2n}$&$q_{k,3n}$&$q_{k,4n}$ &
		$q_{k,1d}$&$q_{k,2d}$&$q_{k,3d}$&$q_{k,4d}$ \\
		\hline
		1& -0.04047945 & 0.11134466 & -0.12450449 & 0.06478750 & -0.01337327 & -2.93860831 & 3.79902478 & -2.61981879& 0.84469060 \\
		3& 3.4415e-04 & -0.00112178 & 0.00120661 & -3.5765e-04 & -8.2661e-05 & -3.73942745 & 5.37449512 & -3.55442936 & 0.92870411 \\
		5& -3.8228e-05 & 1.5320e-04 & -2.5454e-04 & 2.0114e-04 & -6.0832e-05 & -2.29784880 & 1.47464250 & 0.0 & -0.17185948  \\
		\hline
		\hline
	\end{tabular}
	\caption{The parameter values can be utilized to build the $\hmu_B$ dependent part of the equation of state.}
	\label{tab:pfit}
\end{table}

\begin{eqnarray}
N^B_k &=&  
\frac{N^B_{k,0n} + N^B_{k,1n} \bar{t}+N^B_{k,2n} \bar{t}^2+
N^B_{k,3n} \bar{t}^3+N^B_{k,4n} \bar{t}^4}{1+N^B_{k,1d} \bar{t}+
N^B_{k,2d} \bar{t}^2+
N^B_{k,3d} \bar{t}^3+N^B_{k,4d} \bar{t}^4} \;\; , \;\; k=1,\ 3,\ 5 \;\; . \\
q_k &=&  
\frac{q_{k,0n}+ q_{k,1n} \bar{t}+q_{k,2n} \bar{t}^2+
q_{k,3n} \bar{t}^3+q_{k,4n} \bar{t}^4}{1+q_{k,1d} \bar{t}+q_{k,2d} \bar{t}^2+
q_{k,3d} \bar{t}^3+q_{k,4d} \bar{t}^4} \;\; , \;\; k=1,\ 3,\ 5 \;\; .
\label{NBfit}
\end{eqnarray}
Here $\bar{t} = T_0/T$ with an arbitrary temperature scale
$T_0=154$~MeV used as a normalization. The values of the parameters used in these functions are listed in Table \ref{tab:pfit}. The temperature derivatives can be obtained by taking the analytical derivatives of these functions. The consistency of this parametrization is tested on the expansion coefficients shown in Fig.~\ref{fig:exp_coeff}. By combining the parametrization found in Table \ref{tab:pfit} with the $\mu_B = 0$ parametrization found in \cite{HotQCD:2014kol}, we have produced the central lines shown in Fig.~\ref{fig:pes-fixedmu}. The provided parametrization for $p,\epsilon,s$ at non-zero chemical potential accurately describes the lattice QCD data in the temperature range $T \in [135:280]$ and chemical potential $\mu_B / T \in [0:2.5]$. These parametrization can also be used to calculate speed of sound and other transport coefficients.

In heavy ion experiments, strongly interacting matter is created when nuclei collide and then its expand and 
cool while following lines of constant entropy per net baryon number. Thus, to obtain an EoS for fixed $\{n_S=0,n_Q/n_B=0.4,s/n_B\}$
we must solve Eq.~\ref{eq:s_nB} for $\mu_B/T$ to determine 
the $\mu_B/T$ vs. $T$ trajectories that keep $s/n_B$ fixed \cite{david}.
\begin{eqnarray}
	\frac{s}{n_B} = \frac{\sigma_0 + \sum\limits_{k=1}^{\infty} \sigma_{2k}(T) \hmu_B^{2k} }{\sum\limits_{k=1}^{\infty} N_{2k-1}(T) \hmu_B^{2k-1}}
	\label{eq:s_nB}
\end{eqnarray}
 In Figure \ref{fig:pes-fxiedsB} (left), we plot the $\mu_B/T$ against temperature ($T$) for various values of $s/n_B$. It can be observed that $\mu_B / T$ remains relatively constant at temperatures above 250 MeV for fixed $s/n_B$ values greater than 100. However, for smaller $s/n_B$ values, $\mu_B / T$ increases at both high and low temperatures. We also compare these curves to those of ideal gas curves and find that the deviation is approximately 20\% for all $s/n_B$ values at the highest temperature. Additionally, we used the previously mentioned parametrization to calculate these trajectories and plotted them as black solid lines.
 
 In Fig.~\ref{fig:pes-fxiedsB} (right), we present the temperature dependence of pressure, energy, and entropy densities at fixed $s/n_B$ trajectories. These observables exhibit similar behavior to those obtained at fixed $\hmu_B$, as shown in Fig.~\ref{fig:pes-fixedmu}. However, above the pseudo-critical temperature, we observe a sharp increase in these observables for smaller $s/n_B$ as the $\mu_B/T$ increases at high temperatures. We also include solid lines representing the parametrization of $p$, $\epsilon$, and $s$ on fixed $s/n_B$ trajectories in these figures. In Table \ref{tab:eos}, we provide some preliminary numbers for the pressure,energy and entropy density near the pseudo-critical temperature $\Tpc(T,\mu_B/T) = T_{\text{pc},0}(1 - \kappa_2(\mu_B/T)^2)$ \cite{HotQCD:2018pds}. The EoS for fixed $s/n_B$ calculated here is applicable to the range of beam energies currently accessible by BES-II at RHIC in collider mode, which is $7.7~{\rm GeV}\le \sqrt{s_{_{NN}}} \le 200~{\rm GeV}$.

 \begin{table}[t]
 	\centering
 	\begin{tabular}{|c||c|c|c|c|c|} \hline\hline
 		$s/n_B$&$\Tpc$ [MeV]&$\mu_B/T$ & $p$ [MeV/$\text{fm}^3$] &$\epsilon$   [MeV/$\text{fm}^3$] &s   [MeV/$\text{fm}^2$] \\
 		\hline
 		400  & 155 & 0.15(2) & 55(7) & 355(42) & 512(60) \\
 		50  & 150 & 1.2(1)  & 47(6) & 301(35) & 446(52) \\
 		30  &  145 &  2.5(6) & 48(9) & 336(83) & 525(107) \\
 		\hline
 		\hline
 		\end{tabular}
 	\caption{Value of pressure, energy and entropy density for different $s/n_B$ and correspoding $T$ and $\mu_B / T$.}
 	\label{tab:eos}
 \end{table}

\section{Summary}\label{sec:outlook}
In this study, we focused on the Taylor expansion of various thermodynamic quantities, 
such as pressure, energy, and entropy densities, for strange neutral matter with a electric charge to baryon number density ratio of $r=0.4$. 
We observed that $\hmu_B$ dependent part of these observables converge faster 
to their ideal gas values at high temperatures. We also developed the equation of state for this matter 
and found that it can be accurately described using \pade approximants. Our comparison of Taylor expansions 
and \pade approximants at certain $\hmu_B$ values has given us confidence in the validity of our 
Taylor expansion results within a range that varies from $\hmu_B \simeq 2.5$ at low temperatures to $\hmu_B \gtrsim 3$ at temperatures above $200$ MeV. 
Additionally, we have updated the parametrization of the EoS for the $\mu_B$-dependent part. 
We also present the EoS of (2+1)-flavor QCD on the fixed $s/n_B$ trajectories relevant for BES II at RHIC. 

In the future, we aim to smooth out the "wiggles" observed in energy and entropy densities at $\mu_B / T = 2.5$ by constructing \pade approximants for these quantities using the pressure \pade and utilizing thermodynamic relations. This method has been successfully applied for the vanishing electric charge chemical potential in \cite{HotQCD_Eos}.
\section*{Acknowledgements}
This work was supported by the DFG Collaborative Research Centre
315477589-TRR 211, ``Strong interaction matter under extreme conditions”. We thank all the members of the HotQCD collaboration for very helpful discussions.
%
%

\bibliographystyle{JHEP}
\bibliography{bibliography}

\end{document}